\documentclass[dvips]{article}
\usepackage{amsmath,amssymb}
\usepackage{latexsym}
\usepackage{afterpage}
\usepackage{icrctc07}

\title{Calculation of the atmospheric muon flux motivated by the ATIC-2 experiment}
\shorttitle{Calculation of the atmospheric muon flux motivated by the ATIC-2 experiment}

\authors{A.~A.~Kochanov$^{1}$, A.~D.~Panov$^{2}$, T.~S.~Sinegovskaya$^{1}$ and
S.~I.~Sinegovsky$^{1}$.}

\shortauthors{A.~A.~Kochanov, A.~D.~Panov, T.~S.~Sinegovskaya, 
S.~I.~Sinegovsky.}

\afiliations{$^1$Irkutsk State University, Irkutsk, Russia\\ 
             $^2$Skobeltsyn Institute of Nuclear Physics, Moscow State University, Moscow,   Russia}
\email{panov@dec1.sinp.msu.ru}

\abstract{We calculate the cosmic ray muon flux at ground level using directly the primary cosmic ray spectrum and composition measured  in the ATIC-2 balloon experiment. In order to extend the calculations to more high energies, up to $100$ TeV, we use the data of the  GAMMA experiment as well as Zatsepin and Sokolskaya model. This model supported by the ATIC-2 data comprises contributions to the cosmic ray flux of three classes of astrophysical sources -- the shocks from exploding stars, nova and supernova of different types.  The muon flux computation is based on the method for solution of atmospheric hadron cascade equations in which rising total inelastic cross-sections of hadron-nuclear interactions as well as non-power-law character of the primary cosmic ray spectrum are taken into account. The calculated  muon spectrum agrees well with measurements of L3+Cosmic and BESS-TeV, CAPRICE, Frejus, MACRO, LVD as well as other experiments.}

\begin{document}
\maketitle

\section{Introduction}

The muons produced through the cosmic ray interactions with the Earth atmosphere 
provide the tool for indirect study of the primary cosmic ray (PCR) spectra.
May comparison of the predicted and measured atmospheric muon (AM) flux serve as reliability trial for PCR data? The answer depends on the relationship between size of the PCR uncertainties and that of AM flux.  To attempt answering the question we calculate the cosmic ray muon flux at the ground level using directly the data on PCR spectrum and composition measured in the ATIC-2 experiment~\cite{ATIC2}.

 In order to compare the predictions with the high-energy measurements of the AM flux we extend the calculations to more high energies, up to $100$ TeV, using also the PCR spectrum data of the GAMMA experiment~\cite{GAMMA}.
 The PCR model by Zatsepin and Sokolskaya~\cite{ZS3CM} supported by the ATIC-2 data was applied as the nice instrument to extrapolate median energy data to high energy one. 
This model comprises contributions to the cosmic ray flux of three classes of astrophysical sources like supernova and nova blast waves (shocks).
 The muon flux calculation is based on the method to solve the atmospheric hadron cascade equations~\cite{NS, KCC} in which we take into account rising total inelastic cross-sections of hadron-nuclear interactions as well as non-power-law character of the primary cosmic ray spectrum.  A high convergence of the method provides an operative way to calculate the secondary cosmic ray fluxes and allows to test "on-the-fly" the  primary spectrum models. 
 
Hadron fluxes were computed with slightly revised Kimel and Mokhov parametrization (see ~\cite{KMN, NS}) for  nucleon and meson production cross sections which are close to the SIBYLL mini-jet model~\cite{SIBYLL}.

\section{Primary cosmic ray spectra}

\begin{figure*}[ht]
	\centering
\includegraphics[width=0.83\textwidth, trim = 0cm 0.6cm 0cm 0.7cm]{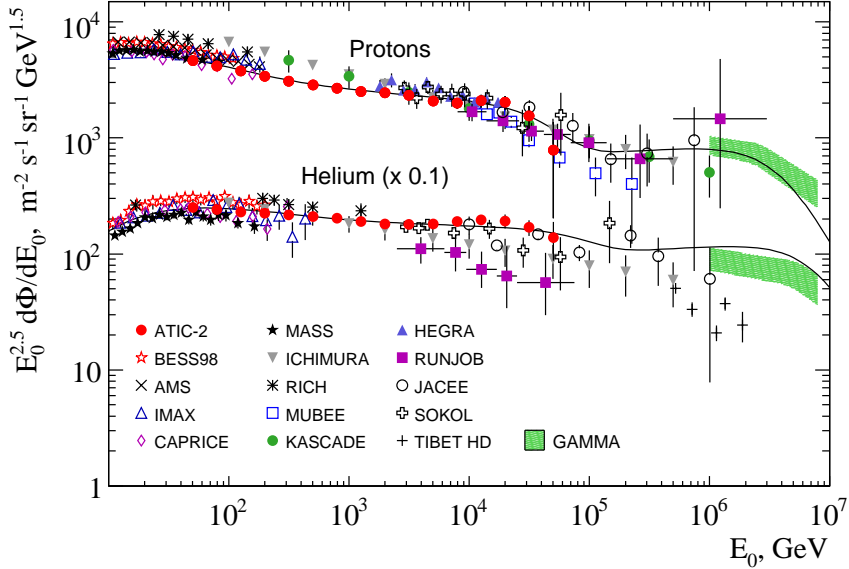}
	\caption{Primary proton and helium spectra, combining
  balloon, satellite and ground-based measurements. The solid curve presents Zatsepin and Sokolskaya model~\cite{ZS3CM}. $E_0$ is the kinetic energy of the particle.}  
		\label{PCR}
\end{figure*}
The balloon borne experiment ATIC (Advanced Thin Ionization Calorimeter)~\cite{ATIC2}, designed for measurements of cosmic rays energy spectra with individual charge resolution from protons to iron, enabled to obtain PCR spectra in the wide energy interval $50$ GeV -- $200$ TeV  with high statistical assurance. 
The differential spectra of protons and helium nuclei obtained in the ATIC-2 experiment are shown in figure~\ref{PCR} along with a bulk of data from balloon, satellite and ground based experiments -- BESS~\cite{BESS98}, AMS~\cite{AMS}, 
IMAX~\cite{IMAX}, CAPRICE~\cite{CAP}, MASS~\cite{MASS}, RICH~\cite{RICH}, MUBEE~\cite{MUBEE},  RUNJOB~\cite{RUNJOB}, JACEE~\cite{JACEE}, SOKOL~\cite{SOKOL},  KASCADE SH~\cite{KASSH}, GAMMA~\cite{GAMMA}, HEGRA~\cite{HEGRA}, TIBET HD~\cite{TIBETHD}, ICHIMURA~\cite{ICHIMURA}. 

Proton and helium spectra measured in the ATIC-2 experiment have different slopes and differ from a simple power law. 
The ATIC-2 data are in agreement with the data of magnetic spectrometers (BESS, AMS, IMAX, CAPRICE, MASS) below $100$ GeV. In the energy region $1<E<10$ TeV the ATIC-2 data are consistent with the SOKOL  measurements and  with those of atmospheric Cherenkov light detector HEGRA. At energies above $\sim 10$ TeV the spectra become steeper, and follow the data of emulsion chamber experiments MUBEE and JACEE, though the agreement is not so clear.
The solid curves in figure~\ref{PCR} are to present the model suggested by Zatsepin and Sokolskaya (ZS)~\cite{ZS3CM} that fits well the ATIC-2 experimental data and describe PCR spectra in the energy range $10$--$10^7$ GeV. 
In order to extend our calculation to higher energies, the PCR spectra measured in the GAMMA~\cite{GAMMA} experiment  was used.
The energy spectra and elemental composition, obtained in the GAMMA experiment cover the $10^3$--$10^5$ TeV range (shaded areas) and agree with the corresponding extrapolations of known balloon and satellite data at the $E\geq10^3$ TeV. In the present calculation, a version of the spectra, reconstructed in the framework of $1,2$D combined analysis with the SIBYLL interaction model (see~\cite{GAMMA} for details), was utilized.

\section{Conventional atmospheric muons} 

\afterpage{\clearpage}
\begin{figure*}[ht]
	\begin{center}
\includegraphics[width=0.65\textwidth, trim = 1cm 0.7cm 1cm 0cm]{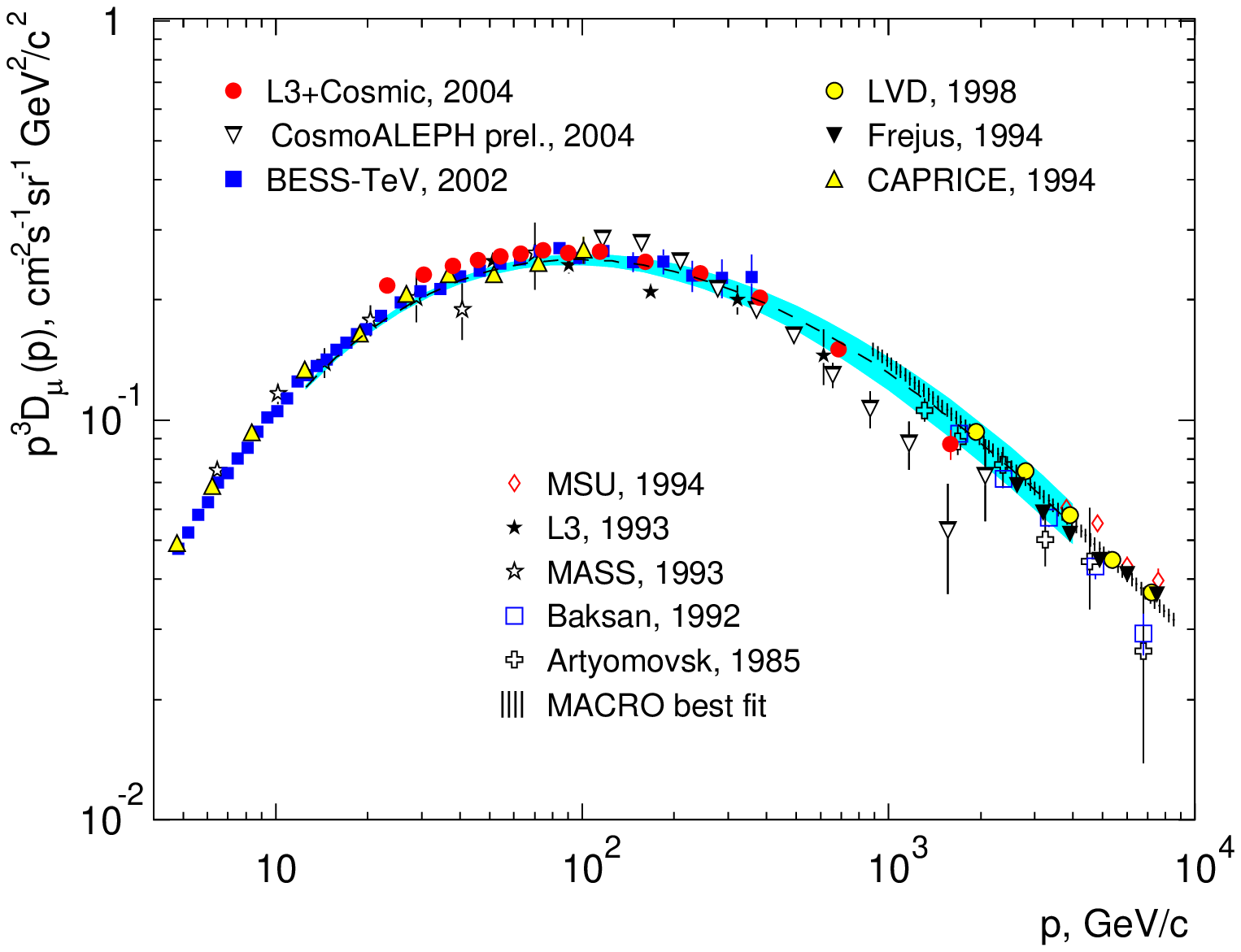}
	\end{center}
\caption{Energy spectrum of muons at ground level near vertical. The dashed-line curves and the shaded area present this work calculation with the ATIC-2 primary cosmic-ray spectrum.}
\vskip 1 cm
	\label{mu-1}
	\begin{center}
\includegraphics[width=0.65\textwidth, trim = 1.15cm 0.7cm 1cm 0cm]{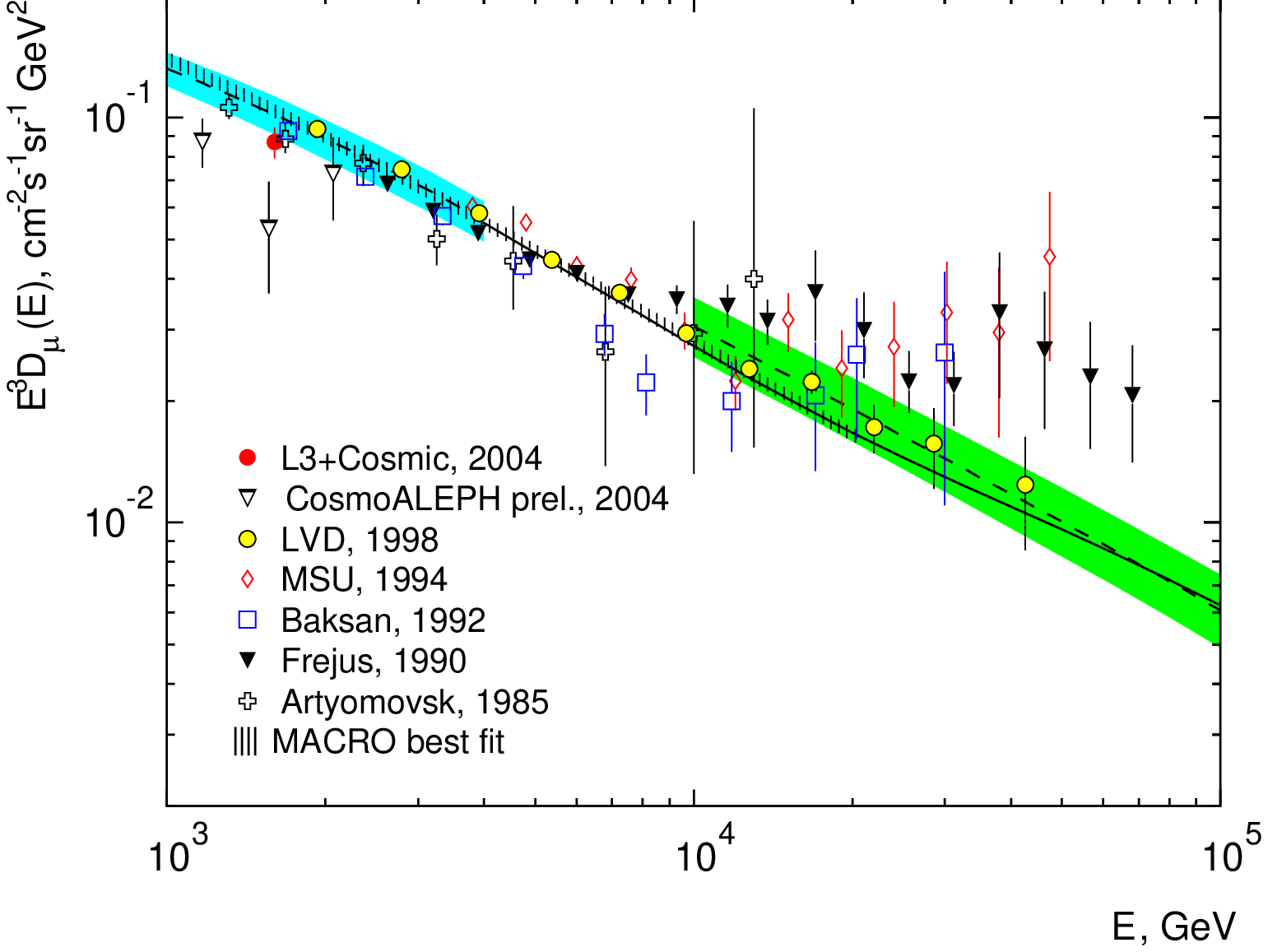}
	\end{center}
	\caption{High-energy plot of the ground level muon spectrum. The dashed-line curves and shaded areas present this work calculations with the ATIC-2 primary spectrum (left) and GAMMA one (right). The solid curve presents the calculation with Zatsepin and Sokolskaya model.}
\label{mu-2}
\end{figure*}
Apart from evident sources of AM, $\pi_{\mu2}$ and  $K_{\mu2}$ decays, 
we take into consideration three-particle semileptonic decays, $K^{\pm}_{\mu3}$, $K^{0}_{\mu3}$. 
Also we take into account small fraction of the muon flux originated  from decay chains   $K\rightarrow\pi\rightarrow\mu$ ~  ($K^0_S\rightarrow \pi^+ + \pi^-$,
$K^\pm \rightarrow \pi^\pm +\pi ^0$, $K^0_{L}\rightarrow \pi^\pm+
\ell^\mp+ \bar{\nu}_{\ell}(\nu_{\ell})$, $\ell=e, \mu $).
We do not consider here a conjectural prompt muon component of the flux (see e.g. \cite{Bugaev98, Misaki03}).
 In figures~\ref{mu-1}, \ref{mu-2} presented are results of the calculation of the surface muon flux along with the data of muon experiments that comprise the direct
measurements of CAPRICE~\cite{CAPRICE94}, BESS-TeV \cite{BESSTEV}, L3+Cosmic~\cite{L3C}, Cosmo-ALEPH (see Ref.~\cite{Lecoultre}), L3 and MASS (taken from~\cite{Bugaev98}) as well as the data (converted to the surface) of underground experiments MSU~\cite{MSU}, MACRO \cite{MACRO}, LVD~\cite{LVD}, Frejus~\cite{Frejus}, Baksan~\cite{Baksan}, Artyomovsk~\cite{ASD}. 
The light shaded areas in figure~\ref{mu-1} and figure~\ref{mu-2} (the left corner)  show the muon spectrum calculated with the ATIC-2 primary spectra taking into consideration statistical errors (dashed curve corresponds to mean values). For the range $10-3000$ GeV one sees fair accordance of the muon flux, calculated with the ATIC-2 spectra, and the recent measurements  but the Cosmo-ALEPH data.
 The high-energy part of the muon flux is shown in figure~\ref{mu-2}, where the dark shaded area (at the right) presents our calculation with the GAMMA primary spectra input and the solid curve presents the muon flux computed with ZS primary spectrum model which appears to be a reliable bridge from TeV range to PeV one. 
It should be noted that without considering the prompt muon contribution above $10$ TeV one can say about satisfactory agreement of calculated fluxes only with the data of MACRO and LVD measurements.

In conclusion, it may be said that the high accuracy of the ATIC-2 data results in the muon flux calculation uncertainty, comparable with rather high precision of the last decade muon flux measurements. 

\section{Acknowledgments}
We are grateful to V.~I.~Zatsepin and S.~V.~Ter-Antonyan for useful discussions.
This work was supported  by Federal Programme  "Leading Scientific Schools of the Russian Federation", grant NSh-5362.2006.2.

\end{document}